\documentclass[11pt]{article}
\usepackage{xspace}
\usepackage{graphicx}
\usepackage{amsmath}
\usepackage{amssymb}
\usepackage{color}
\usepackage{units}
\usepackage{units}

\definecolor{Red}{rgb}{1,0,0}
\definecolor{Green}{rgb}{0,1,0}
\definecolor{Blue}{rgb}{0,0,1}
\definecolor{Black}{rgb}{0,0,0}

\def\beq{\begin{equation}}
\def\eeq#1{\label{#1}\end{equation}}
\def\eeqn{\end{equation}}

\def\beqa{\begin{eqnarray}}
\def\eeqa#1{\label{#1}\end{eqnarray}}
\def\eeqan{\end{eqnarray}}

\def\overbar#1{\overline{#1}}

\let\bar=\overbar

\def\Dslash{\not{\hbox{\kern-4pt $D$}}}
\def\dslash{\not{\hbox{\kern-2pt $\del$}}}

\def\msb{{\bar{\ssstyle M \kern -1pt S}}}

\newcommand\knnb {\ensuremath{ K^{+} \rightarrow \pi^{+} \nu \overbar{ \nu }  } }
\newcommand{\mmpi}{m^{2}_{\textrm{miss}}}

\def\Title#1{\begin{center} {\Large {\bf #1} } \end{center}}

\begin{document}

\Title{Prospects for \knnb
observation at CERN in NA62 }

\bigskip\bigskip


\begin{raggedright}  

Francis Newson\index{Newson, F.}, {\it University of Birmingham}\\

\begin{center}\emph{for the NA62 Collaboration 
\footnote{G.~Aglieri Rinella, R.~Aliberti, F.~Ambrosino, B.~Angelucci, A.~Antonelli, G.~Anzivino, 
R.~Arcidiacono, I.~Azhinenko, S.~Balev, J.~Bendotti
% ...
}.
}
\end{center}

\bigskip
\end{raggedright}

{\small
\begin{flushleft}
\emph{The NA62 experiment will begin taking data in 2015. 
    Its primary purpose is a 10\% measurement of the branching ratio
    of the ultrarare kaon decay $K^{+} \rightarrow \pi^{+} \nu \overbar{\nu}$, using
the decay in flight of kaons in an unseparated  beam with momentum \unit[75]{GeV/c}.
The detector and analysis technique are described here. }
\end{flushleft}
}

\section{Introduction}
The NA62 experiment is a kaon decay in flight experiment at the CERN SPS.
Its primary aim is to measure the branching ratio of the ultra-rare decay $\knnb$ with 10\% precision by collecting $\mathcal{O}(100)$ signal events\cite{Anelli:2005ju}.

The decay $\knnb$ is a Flavour Changing Neutral Current process so, in the Standard Model, it is forbidden at tree level and proceeds through box and penguin diagrams.
The hadronic matrix element can be determined from the (isospin rotated) semileptonic decay $K^{+} \rightarrow e^{+} \nu_{e} \pi^{0}$, which is well measured \cite{Agashe:2014kda}.
The electroweak amplitude is largely dominated by top-quark loops, resulting in a theoretically clean dependence on the product of CKM matrix elements, $V^{*}_{ts}V_{td}$.
The small branching ratio and CKM suppression make this decay a sensitive probe of new physics, especially in non Minimal Flavour Violation models\cite{Isidori:2006qy,Blanke:2009am}.
Incorporating two-loop electroweak corrections\cite{Brod:2011feb}  has reduced the theoretical uncertainty on the branching ratio, so the Standard Model estimate is now limited by parametric errors. The current expected value is $\mathcal{B}( \knnb) = ( 7.81^{+0.80}_{-0.71}\pm0.29)\times 10 ^{-11}$. (Here, the first error is parametric and the second is the purely theoretical uncertainty).

The current experimental value comes from the E787 and E949 experiments at the Brookhaven National Laboratory (BNL). Both experiments identified $\knnb$ events by detecting the outgoing pion from kaons decays at rest. 
They recorded a combined total of 7 events, yielding a branching ratio measurement $\mathcal{B}( \knnb) = 1.73^{+1.15}_{-1.05} \times 10 ^{-10}$\cite{Artamonov:2008qb}.

\section{Measurement principle}
A measurement with 10\% precision will require detecting $\mathcal{O}(100)$ events while controlling systematic errors at the percent level\cite{Anelli:2005ju}.

Large statistics can be achieved by using a high intensity kaon beam and maximizing signal acceptance.
The NA62 experiment \cite{Hahn:2010dec} will use protons from the SPS beam to produce a secondary beam with momentum $\unit[75]{GeV/c^2}$, consisting of kaons, protons and pions.
The total instantaneous beam rate is $\unit[750]{MHz}$, resulting in $4.5\times10^{12}$ $K^{+}$ decays per year.

\begin{table}[!th]
\begin{center}
    \caption{The most common $K^{+}$ decay modes along with the strategies for suppressing them at NA62.}
\begin{tabular}{lcccc}  \hline\hline
Decay Channel &  Branching ratio (\%) &  Suppression Strategy  \\
    \hline
    $K^{+} \rightarrow \mu^{+} \nu$              &  $63.55  \pm 0.11 $  &  $\mu$ veto + 2-body kinematics\\
    $K^{+} \rightarrow \pi^{+} \pi^{0}$          &  $20.66  \pm 0.08 $  &  Photon veto + 2-body kinematics\\
    $K^{+} \rightarrow \pi^{+} \pi^{+} \pi^{-}$  &  $ 5.59  \pm 0.04 $  &  Charged particle veto + kinematics\\
    $K^{+} \rightarrow \pi^{0} e^{+} \nu $       &  $ 5.07  \pm 0.04 $  &  $E/p$ + photon veto\\
    $K^{+} \rightarrow \pi^{0} \mu^{+} \nu $     &  $ 3.353 \pm 0.034$  &  $\mu$ veto + photon veto\\
    $K^{+} \rightarrow \pi^{+} \pi^{0} \pi^{0}$  &  $ 1.761 \pm 0.022$  &  Photon veto + kinematics\\
    \hline\hline
\end{tabular}
\label{tab:backgrounds}
\end{center}
\end{table}

Systematic errors are controlled with large background rejection and high redundancy between subdetectors.
The signal signature is a single $K^{+}$ upstream matched with a single $\pi^+$ track downstream, and no other particles detected.
Most backgrounds to the signal decay come from other kaon decays with similar decay signatures when one or more of the decay products is mis-identified or not detected.
The most common decays are shown in table \ref{tab:backgrounds} \cite{Anelli:2005ju, Agashe:2014kda}.
The experimental method combines high resolution particle tracking and momentum measurment with particle identification and vetoing to achieve a signal-background ratio $S/B \approx 10$.

Another potential background comes from beam induced accidentals - a scattered pion from the beam has the same experimental signature as the signal decay.
Accidentals of this kind can be suppressed by tagging the kaon before it decays.
Precise timing is required to match the tagged kaon to its decay products downstream.

\begin{figure}[!ht]
    \begin{center}
        \includegraphics[width=0.6\columnwidth]{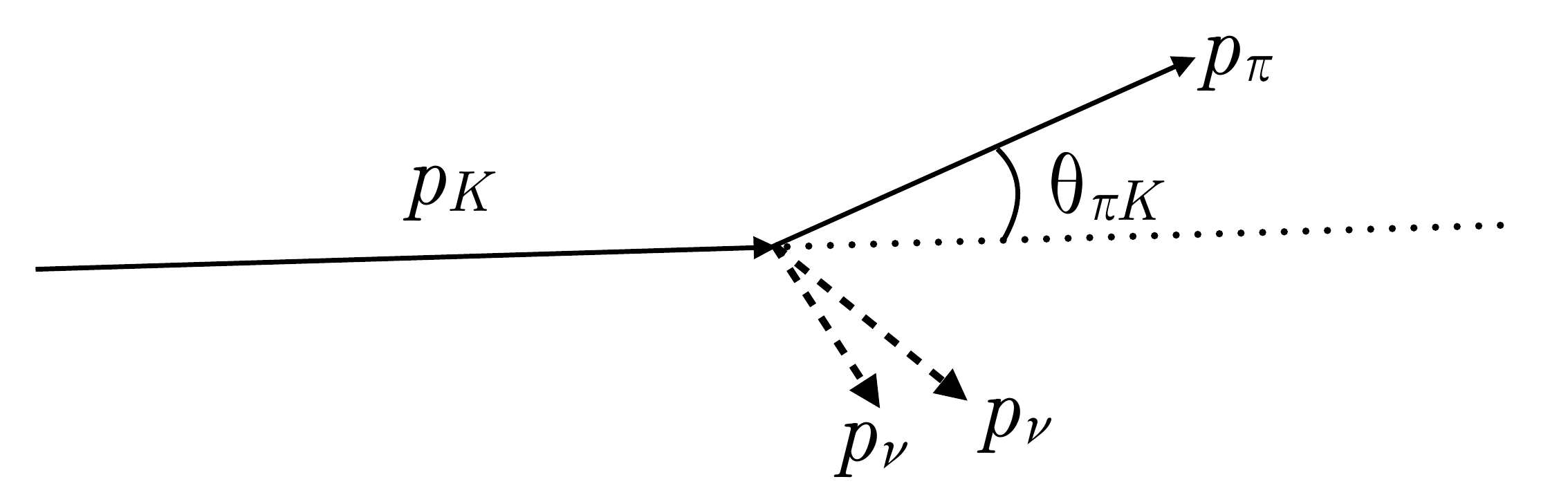}
        \caption{Kinematics of the $\knnb$ decay}
        \label{fig:kinematics}
    \end{center}
\end{figure}

Backgrounds can be described in terms of their kinematic distribution.
Figure \ref{fig:kinematics}, shows a sketch of the kinematics for $\knnb$.
Since the two neutrinos are undetectable, only the kaon and pion three-momenta can be measured.
From these, one can compute the squared missing mass in the pion hypothesis:
\begin{align}
    \mmpi &\equiv (p_K - p_{\pi})^2 \\
    &= m^2_K  + m^2_{\pi} - 2E_K E_{\pi} + 2 |\vec{p}_{K}||\vec{p}_{\pi}| cos \theta_{\pi K}
\end{align}
Here $p_K$,$p_\pi$ are the particle four-momenta, $\vec{p_K},\vec{p_\pi}$ are the three-momenta and the energies $E_K,E_\pi$ are computed using $E=\sqrt{\vec{p}^2 + m^2}$. $\theta_{\pi K}$ is the angle between the pion and kaon three-momenta.

\begin{figure}[!ht]
    \begin{center}
        \includegraphics[width=0.65\columnwidth]{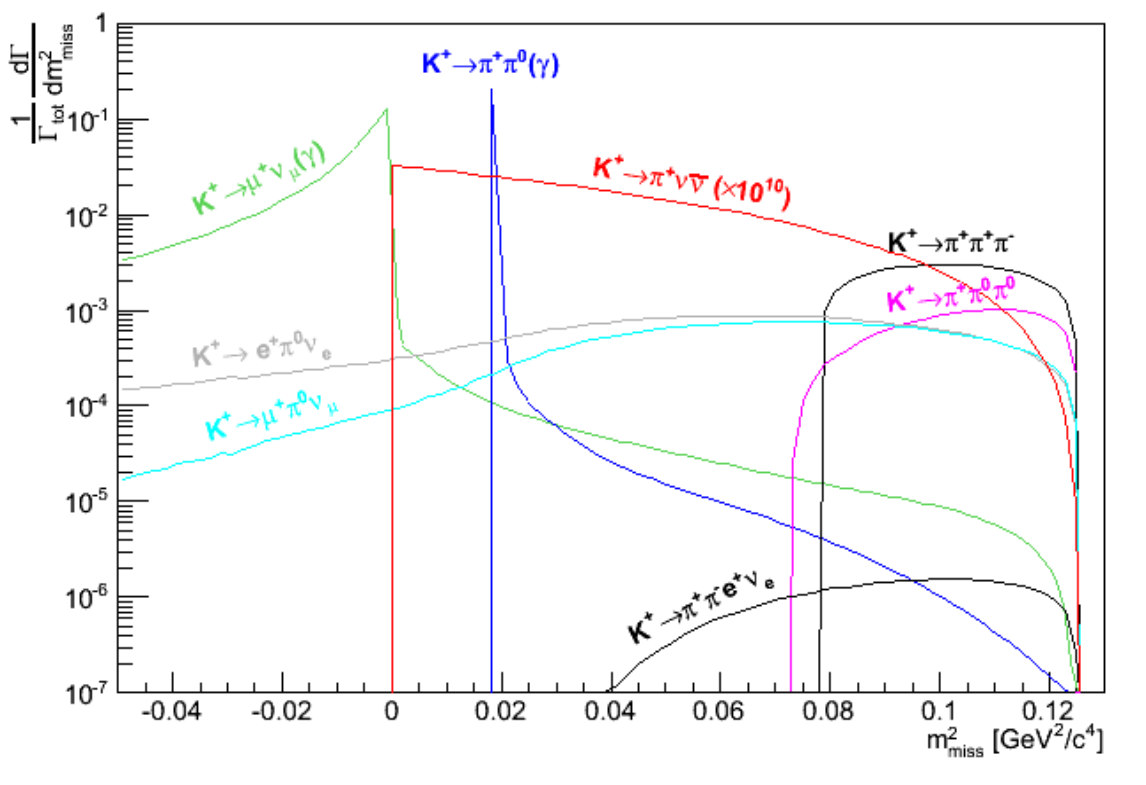}
        \caption{Distribution of $m_\textrm{miss}^2$ for the main background channels.
        The signal distribution is mulitplied by a factor of $10^{10}$}
        \label{fig:bgkinematics}
    \end{center}
\end{figure}

Figure \ref{fig:bgkinematics} shows the distribution of $\mmpi$ for the main backgrounds and the signal \cite{Ruggiero:2013oxa}.
The signal decay has a continuous spectrum due to the unmeasured neutrinos, while the largest background, $K^{+}\rightarrow \pi^+ \pi^{0}$, is peaked at ${\mmpi=m^2_{\pi^{0}}}$.
At higher values of $\mmpi$, there is a sharp turn on for the decay $K^{+} \rightarrow \pi^{+}\pi^{+}\pi^{-}$, while at negative missing mass there is a large contribution from $K^{+}\rightarrow \mu^{+} \nu$ due to the wrong mass hypothesis.
These three backgrounds naturally define signal regions where a minimum background is expected, away from the $K^{+}\rightarrow \pi^{+} \pi^{0}$  peak and the $K^{+} \rightarrow \pi^{+}\pi^{+}\pi^{-}$ threshold:
\begin{description}
    \item{Region I:} between 0 and the $K^{+}\rightarrow \pi^{+} \pi^{0}$ peak
    \item{Region II:} between the $K^{+}\rightarrow \pi^{+} \pi^{0}$ peak and the $K^{+} \rightarrow \pi^{+}\pi^{+}\pi^{-}$ threshold $\approx 4 m^{2}_{\pi}$
\end{description}
Restricting the allowed kinematic space in this way removes 92\% of the background. The remaining background must be suppressed by using photon and muon vetos to detect extra particles in the final state and using particle identification to distinguish between pions and other particles (i.e. muons and electrons).

The main background in the allowed kinematic regions comes from $K{^+}\rightarrow \pi^{+}\pi^{0}$ decays with an additional radiated photon.
It is important to efficiently detect the $\pi^{0}$ in order to reject these events.
Here the decay in flight technique has an advantage with respect to techniques using stopped kaons since the initial kaon momentum of $\unit[75]{GeV/c}$ means that the decay products are boosted in the lab frame so they can be detected more efficiently.
In particular, requiring $|\vec{p}_{\pi^{+}}| < \unit[35]{GeV/c} $ ensures that the photons from the associated $\pi_{0}$ in $K^{+} \rightarrow \pi^{+} \pi^{0}$ have a total energy of $~\unit[40]{GeV}$.

\section{The NA62 Detector}
\begin{figure}[!ht]
    \begin{center}
        \includegraphics[width=\columnwidth,height=0.3\columnwidth]{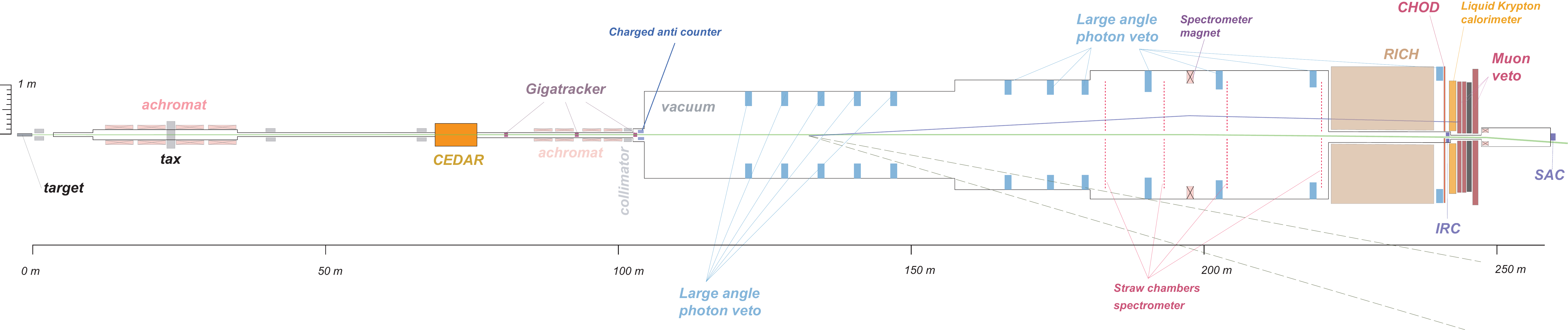}
        \caption{The NA62 Experimental Setup}
        \label{fig:layout}
    \end{center}
\end{figure}
Figure \ref{fig:layout} shows the layout of the NA62 detector, which will be placed on the same CERN-SPS extraction line used by NA62 in 2007 for the measurment of $R_K$\cite{Lazzeroni:2012cx}.
\subsection{Beam}
The NA62 experiment will use an unseparated beam, obtained by impinging $\unit[400]{GeV/c}$ protons from the SPS onto a beryllium target.
The secondary beam is selected to have a momentum $\unit[75]{GeV/c}$, with 1\% momentum bite.
At the entrance to the CEDAR detector, upstream of the decay volume, it consists of approximately 6\% kaons, 22\% protons and 72\% pions.
Beam particles are allowed to decay in a \unit[65]{m} long decay vacuum which is evacuated (to $< \unit[10^{-6}]{mbar}$) in order to reduce backgrounds from scattering.

\subsection{Tracking and momentum measurement systems}
It is vital to have accurate measurement of both the kaon and pion momenta in order to use the reconstructed missing mass as a selection variable. 
In order to suppress the tails of backgrounds on the edge of the allowed kinematic region, the resolution on $\mmpi$ should be less than $\unit[10^{-3}]{GeV^2/c^2}$.

The kaon spectrometer, called the Gigatracker (GTK) is formed by 3 silicon pixel detectors with the same dimensions as the beam. They are placed before, in the centre of, and after an achromat upstream of the decay volume in order to measure the kaon momentum  with a relative resolution of $\sigma(p_{K})/p_{K} \approx 0.2\%$ and the direction with a resolution of the order of $\unit[16]{\mu rad}$.
This resolution can be achieved with pixels of size $\unit[300]{\mu m} \times \unit[300]{\mu m}$ and a sensor thickness of $\unit[200]{\mu m}$.
The low material budget is required not only to achieve good resolution in GTK but also to prevent scattered particles from interacting with other detectors downstream.

Since the GTK is exposed to a high flux of particles, it must have a good time resolution in order to correctly match kaons with their decay products downstream.
The readout system ensures a time resolution of better than \unit[200]{ps} for each pixel hit, resulting in a track resolution better than \unit[150]{ps}.

Even with the ultra thin GTK detectors, there is a background induced by inelastic interactions of the beam with the final GTK station.
If a pion is emitted at low angle, it can reach the STRAW tracker and mimic a signal event.
The Charged ANTI (CHANTI) detector is designed to veto inelastic interactions in the GTK by detecting particles emitted between \unit[34]{mrad} and \unit[1.38]{rad} with respect to the beam axis.
It consists of 6 stations, placed downstream of the GTK, each consisting of two layers of scintillating bars placed at $90^{\circ}$ to each other.
For signal like events, the CHIANTI vetoes events with inelastic collisions with almost 99\% efficiency.

Kaon decay product momenta are measured by a magnetic spectrometer, the Straw Tracker, which consists of four chambers in vacuum, two upstream and two downstream of a dipole magnet producing a vertical B-field of \unit[0.36]{T}.
Secondary particles must be reconstructed with momentum resolution $\sigma(p)/p \leq 1\%$ and $\sigma(\theta_{K \pi}) \leq \unit[60]{\mu rad }$ with minimal Coulomb scattering, particularly in the first chamber.
These requirements imply a spatial resolution $\leq \unit[130]{\mu m}$ per coordinate and material budget of $\leq 0.5\% $ of a radiation length for each chamber.

Each chamber contains 4 views, measuring coordinates along the $x$, $y$ and $\pm 45^{\circ}$ directions.
Each view consists of 448 straw tubes, arranged in 4 staggered layers, with a central strip left empty to create an octagonal hole for undecayed beam particles to pass through.
The expected performance is $\sigma(|\vec{p}|)/|\vec{p}| \approx 0.3\% \pm 0.007\%$ and $\sigma ( \textrm{d}X, Y / \textrm{d}Z)  \approx \unit[15 - 45]{\mu rad}$, depending on the track momentum.

In addition to the tracking detectors, charged particles can also be detected in the Charged Hodoscope (CHOD).
This is a scintillator hodoscope (reused from the NA48 experiment), with high granularity and good time resolution.
It is used as  a trigger on single charged particles as well as a veto for events with multiple charged particles.

\subsection{Particle identification detectors}
It is important to tag incoming kaons in the beam, because there is a potential background to $\knnb$ from beam pions scattering in the decay volume.
Positively identifying incoming kaons relaxes the requirements on vacuum purity which would otherwise be required.
Pion identification is necessary in order to reject $K^{+} \rightarrow \mu^{+} \nu$ decays at the required level.

Kaon tagging is performed by the CEDAR/KTAG detector. 
CEDAR is a Cherenkov detector built for measuring the composition of SPS beams.
It has been upgraded in order to meet the high rate requirements of NA62.
The expected kaon rate is $\sim \unit[50]{MHz}$, so the new detector will require a timing resolution of better than \unit[100]{ps} on the the kaon crossing time, with a tagging efficiency above 95\%.
Cherenkov radiation emitted from kaons passing though pressurised nitrogen is spread over 384 photomultiplier tubes (PMTs).
On average, $\sim12$ photons are detected per kaon, so the expected rate of photons on individual PMTs is $< \unit[5]{MHz}$.

Muons from the decay $K^{+} \rightarrow \mu^{+} \nu$ are rejected at the $10^5$ level by the Muon Veto system (MUV).
MUV1 and MUV2 are classic iron scintillator sandwich calorimeters consisting of alternating layers of iron and plastic scintillator.
Scintillator strips are alternately oriented horizontally and vertically.
Pions and muons can be distinguished by the shape of their hadronic showers, in conjunction with their electromagnetic showers, measured in the Liquid Krypton detector.
MUV3 is used to detect non-showering muons and act as a muon veto detector at trigger level.
It sits after the other MUV detectors and an \unit[80]{cm} iron wall and consists of a grid of 12 $\times$ 12 scintillator tiles with 8 smaller tiles mounted around the beam pipe.

The RICH is a Ring Imaging Cheronkov detector designed to distinguish between pions and muons to provide an extra factor of $10^2$ suppression for $K^{+} \rightarrow \mu^{+} \nu$ decays. 
It must be able to separate $\pi^{+}$ from $\mu^{+}$ for momenta between 15 and \unit[35]{GeV/c}.
In addition, it provides precise timing for the pion candidate, measuring the crossing time with a resolution better than \unit[100]{ps}.
The detector consists of a tank of neon (at \unit[1]{atm}), \unit[17]{m} long and \unit[3]{m} in diameter. 
Cherenkov light is reflected by a mosaic of mirrors at the downstream end of the detector onto two arrays of about 1000 PMTs each.
The threshold for Cherenkov emission for pions in $|\vec{p_{\pi}|} = \unit[12]{GeV/c}$, so pions are required to have a momentum $ |\vec{p_{\pi}}| \geq  \unit[15]{GeV} $ in order to have enough photons to fit a ring.
Within the momentum range 15 to \unit[35]{GeV/c}, test beam results have shown that $\pi^{+}/\mu^{+}$ separation can be achieved with the requested purity \cite{Angelucci:2010zz}.

\subsection{Photon vetoes}
Many kaon decays can imitate the signal decay if one or more photons escape the detector undetected.
In particular, suppressing $K^{+} \rightarrow \pi^{+} \pi^{0}$ requires $\pi^{0}$ rejection with an inefficiency $< 10^{-8}$ and this has driven the design of the photon veto system.
Hermetic coverage for photons with angles up to \unit[50]{mrad} is divided into three groups.
The Large Angle Vetoes (LAV) cover the region from 8.5 to \unit[50]{mrad}, the Liquid Krypton Electromagnetic calorimeter (LKr) covers the region from 1.5 to \unit[8.5]{mrad} and the Small Angle Vetoes(SAV) cover angles less than \unit[1.5]{mrad}.
Requiring that the $\pi^{+}$ momentum is between 15 and \unit[35]{GeV/c} ensures that the two photons from the $\pi^{0}$ have a total energy of $\sim \unit[40]{GeV}$. A photon is lost out of acceptance in only 0.2\% of $K^{+} \rightarrow \pi^{+} \pi^{0}$ decays and there are no configurations in which both photons are out of acceptance.

The LAV consists of 12 stations made of rings of lead-glass blocks recovered from the OPAL electromagnetic calorimeter barrel\cite{Ahmet:1990eg}.
The first 11 stations intersect the vacuum decay tube, while the last is downstream of the RICH detector.
They have an inefficiency $\sim 10^{-4}$ for photons with energies down to \unit[0.5]{GeV}.

The LKr calorimeter is being reused from the NA48 experiment\cite{Fanti:2007vi}, equipped with a new readout.
It is a quasi-homogeneous ionisation chamber allowing for full development of electromagnetic showers so it helps to provide particle identification as well as acting as a photon veto.
It has an energy resolution $\sigma_{E}/E = 0.032 /\sqrt{E} \oplus 0.09/E \oplus 0.0042$(E in GeV).

The SAV consists of the Small Angle Calorimeter(SAC), placed at the end of the beam line and the Inner Ring Calorimeter (IRC), placed around the beam to cover the angular region between the SAC and the LKr. 
Both are `Shashlyk' detectors based on consecutive lead and plastic scintillator plates.
Light is extracted by wavelength shifting fibres and detected by PMTs.

\subsection{Trigger and data aquisition}
For readout and trigger electronics, most sub-sectors will use a custom board, TEL62 which is derived from the TEL1 board used in LHCb \cite{Sozzi:2010zz}.
A single hardware trigger level (L0) uses information from the RICH, LKr, LAV and MUV to reduce the event rate from $\sim\unit[10]{MHZ}$ to $\sim\unit[1]{MHz}$, while preserving most of the decays of interest.
Following  L0, the L1 software trigger uses information from individual detectors to reduce the rate to $\sim\unit[100]{kHz}$.
Finally the L2 software trigger uses information combined from several detectors to reduce the rate to a few kHZ which can be written to disk.

\section{Summary}
The NA62 experiment has been designed and constructed to meet the requirements for collecting $\mathcal{O}(100)$ events of the $\knnb$ decay.
The data will be collected over approximately two years starting in 2015, after a pilot run at the end of 2014.
The experiment is expected to measure the branching ratio with at least 10\% accuracy, providing a significant test of new physics beyond the Standard Model.

\pagebreak

%
%

%
%
%
%
 
\end{document}